\documentclass[twocolumn,
 amsmath,amssymb,
floatfix]{revtex4-1}
  
\usepackage[left=1.5cm, right=1.5cm, top=1.785cm, bottom=2.0cm]{geometry}
\usepackage{amssymb}
\usepackage{amsmath}
\usepackage{graphicx}
\usepackage{dcolumn}
\usepackage{float}
\usepackage{array}
\usepackage{droidsans}
\usepackage{charter}
\usepackage[T1]{fontenc}
\usepackage[usenames,dvipsnames]{xcolor}
\usepackage{setspace}
\usepackage[compact]{titlesec}
\usepackage{hyperref}
\usepackage{comment}
\usepackage[framemethod=TikZ]{mdframed}
\usepackage{xcolor}
\usepackage{xr}
\usepackage{comment}
\usepackage{listings}
\usepackage{siunitx}
\DeclareSIUnit\torr{Torr}
\definecolor{darkred}{rgb}{0.6, 0.0, 0.0}
\definecolor{darkblue}{rgb}{0.0, 0.0, 0.6}
\definecolor{bg}{rgb}{0.95,0.95,0.95} % Background color for the code
\definecolor{cream}{RGB}{222,217,201}

\definecolor{codebg}{rgb}{0.95, 0.95, 0.95}
\definecolor{codeblue}{rgb}{0.2, 0.2, 0.6}
\definecolor{codegreen}{rgb}{0, 0.6, 0}
\definecolor{codered}{rgb}{0.6, 0, 0}
 
\lstdefinestyle{mystyle}{
    backgroundcolor=\color{codebg},   % Background color
    basicstyle=\ttfamily\footnotesize, % Basic font style
    keywordstyle=\color{codeblue}\bfseries, % Keyword style
    commentstyle=\color{codegreen}\itshape, % Comment style
    stringstyle=\color{codered},      % String style
    numberstyle=\tiny\color{gray},    % Line number style
    stepnumber=1,                     % Step number for line numbers
    numbersep=5pt,                    % Distance between line numbers and code
    showspaces=false,                 % Do not show spaces
    showstringspaces=false,           % Do not show string spaces
    showtabs=false,                   % Do not show tabs
    frame=single,                     % Single line frame
    rulecolor=\color{black},          % Frame color
    framesep=5pt,                     % Frame separation
    framerule=0.5pt,                  % Frame rule width
    breaklines=true,                  % Enable line breaking
    breakatwhitespace=true,           % Break at whitespace
    tabsize=4,                        % Tab size
    captionpos=b,                     % Caption position (bottom)
    keepspaces=true,                  % Keep spaces
    columns=flexible,                 % Flexible column width
}

\begin{document}
\lstset{language=Python}

\title{From Text to Test: AI-Generated Control Software for Materials Science Instruments}
\author{Davi M Fébba}
\email{DaviMarcelo.Febba@nrel.gov}
\author{Kingsley Egbo}
\author{William A. Callahan}
\author{Andriy Zakutayev}
\email{Andriy.Zakutayev@nrel.gov}
\affiliation{Materials Science Center, National Renewable Energy Laboratory (NREL), Golden CO, 80401, USA}

\begin{abstract}
Large language models (LLMs) are one of the AI technologies that are transforming the landscape of chemistry and materials science. Recent examples of LLM-accelerated experimental research include virtual assistants for parsing synthesis recipes from the literature, or using the extracted knowledge to guide synthesis and characterization. However, these AI-driven materials advances are limited to a few laboratories with existing automated instruments and control software, whereas the rest of materials science research remains highly manual. AI-crafted control code for automating scientific instruments would democratize and further accelerate materials research advances, but reports of such AI applications remain scarce. In this work, we demonstrate how to swiftly establish a Python-based control module for a scientific measurement instrument solely through interactions with ChatGPT-4. Through a series of test and correction cycles, we achieved successful management of a common Keithley 2400 electrical source measure unit instrument with minimal human-corrected code. Additionally, a user-friendly graphical user interface (GUI) was created, effectively linking all instrument controls to interactive screen elements. Finally, we integrated this AI-crafted instrument control software with a high-performance stochastic optimization algorithm to facilitate rapid and automated extraction of electronic device parameters related to semiconductor charge transport mechanisms from current-voltage (IV) measurement data. This integration resulted in a comprehensive open-source toolkit for semiconductor device characterization and analysis using IV curve measurements. We demonstrate the application of these tools by acquiring, analyzing and parameterizing IV data from a Pt/Cr\textsubscript{2}O\textsubscript{3}:Mg/$\beta$-Ga\textsubscript{2}O\textsubscript{3} heterojunction diode, a novel stack for high-power and high-temperature electronic devices. This approach underscores the powerful synergy between LLMs and the development of instruments for scientific inquiry, showcasing a path for further acceleration in materials science.
\end{abstract}

\maketitle

\section{Introduction}

Recent advancements in artificial intelligence, particularly the emergence of Large Language Models (LLMs) such as OpenAI's GPT-3 and 4, have sparked the interest of the scientific community. A growing body of knowledge now evaluates LLMs for chemistry and materials science related tasks, such as chemistry knowledge \cite{white_assessment_2023,castro_nascimento_large_2023}, text-parsing of solid state synthesis recipes \cite{thway_harnessing_2024}, domain-specific chatbots for scientific literature \cite{yager_domain-specific_2023}, and also in fully autonomous experimentation workflows \cite{boiko_autonomous_2023}, among many other applications \cite{aal_e_ali_machine_2024}. ChatGPT, specifically, was recently included as a non-human addition to \textit{Nature's 10} due to its profound impact on science in 2023\cite{van_noorden_chatgpt_2023}.

LLMs, endowed with natural language understanding and generation capabilities, have been particularly useful for developing computing code for a myriad of tasks \cite{jablonka_14_2023}, and thus offer unprecedented opportunities for expediting the development of control solutions tailored to laboratory instruments, which traditionally involves intricate programming, extensive testing, and iterative refinement over prolonged periods. However, LLMs revolutionize this process by enabling researchers to articulate desired functionalities in plain language, thereby automating the generation of code snippets or complete control algorithms. Consequently, the development cycle is significantly compressed, allowing researchers to allocate more time to experimental design and scientific inquiry. Furthermore, the agility afforded by LLM-generated code empowers researchers to swiftly adapt control solutions to evolving experimental requirements or unforeseen challenges.

Moreover, the integration of LLMs democratizes control solution development and fosters interdisciplinary collaboration in the scientific community. These models provide a user-friendly interface that simplifies the conversion of ideas into functional code, enabling effective contributions from researchers with diverse backgrounds. Furthermore, applications developed using LLMs, particularly in Python, can amplify the application of the FAIR principles —-- data is Findable, Accessible, Interoperable, and Reusable. Python’s comprehensive suite of open-source libraries inherently makes both the source code and output file formats open source, which significantly contributes to the FAIR data principles by ensuring that data handling practices are transparent and the generated data is readily accessible and usable by the wider research community.

\begin{figure}
    \centering
    \includegraphics[width=\columnwidth]{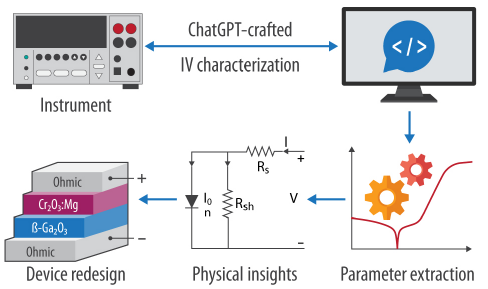}
    \caption{Diagram depicting the AI-engineered current-voltage (IV) characterization module and parameter extraction tool described in this work. The Source Measure Unit (SMU) is connected to a control computer running a Python application developed by ChatGPT. The collected IV characterization data is subsequently processed using a high-performance differential evolution algorithm for automated parameter extraction from the IV curves, to provide insights for redesigning the device stack.}
    \label{fig1}
\end{figure}

The transformative benefits of LLMs in simplifying code development and enhancing data management seamlessly translate to specialized applications across scientific disciplines. Current-voltage (IV) characteristics, for example, play a pivotal role in understanding the fundamental properties and operational behavior of semiconductor devices, serving as a critical tool for diagnosing device performance, efficiency, and for identifying potential areas for optimization. For example, in solar cells, IV curves are essential for determining device-level parameters like open-circuit voltage, short-circuit current, and fill factor, which are key indicators of device efficiency\cite{emery2010}, and also for extracting parameters related to the fundamental physical processes, such as Shockley–Read–Hall recombination parameters\cite{kurchin2020}. Additionally, by modeling the IV curves with equivalent circuits and diode models, parameters such as reverse saturation current, diode ideality factor, series and shunt resistance, can be extracted\cite{ocaya_machine_2023,febba_impacts_2018}. Similarly, in diodes and transistors, IV curves help in assessing threshold voltage, leakage current, and on-off ratio, important parameters for evaluating device switching behavior and power consumption\cite{valdivieso_resistive_2023,zhai_recent_2022,wong_tcad-machine_2020}.

IV measurement and analysis setups are widespread across materials science laboratories, and are often automated by a custom LabView or Python program on a computer.  The automation of these instruments is a task traditionally undertaken by graduate students who invest considerable time in navigating through detailed reference manuals. The analysis step of extracting physical parameters from IV curves also presents a significant challenge due to the nature of the equations resulting from equivalent circuit modeling. Traditional parameter extraction methods often involve tedious, manual adjustments and assumptions that can introduce errors or biases, making the process time-consuming and less accurate. Moreover, the sheer volume of data generated by modern semiconductor testing necessitates a more efficient approach to parameter extraction to keep pace with rapid technological advancements. Metaheuristic optimization algorithms, such as genetic algorithms, simulated annealing, and particle swarm optimization, provide a powerful solution to these challenges and have indeed been successfully applied to this problem in the field of photovoltaics \cite{li_comprehensive_2021}. One example is the Self-Adaptive Differential Evolution\cite{brest_self-adapting_2006}, benchmarked and proven robust against noise on the photovoltaics (PV) parameter extraction problem \cite{febba_effects_2020}. By exploring the vast search spaces associated with IV curve analysis, these algorithms can find optimal or near-optimal parameter values that closely match experimental data, with no need to rely on any model assumptions, making these algorithms compelling alternatives to automate the analysis of IV data.

Here, we demonstrate the application of LLMs, specifically OpenAI's GPT-4 version, to rapidly automate a Keithley 2400 Source Measure Unit (SMU), which is widely used for IV measurements. This automation was accomplished without resorting to model fine-tuning, and did not require extensive programming expertise typically necessitated for such tasks. First, we tasked ChatGPT with generating custom Python code for operating the instrument, and to map the resulting code into a graphical user interface (GUI). The resulting control solution serves as a compelling example of how LLMs can facilitate rapid automation in laboratory settings, especially for researchers with no experience in software development. Furthermore, due to the versatile nature of LLMs, other characterization and synthesis instruments using SCPI, Modbus or other standard commands and protocols, can be automated by the same approach.

Next, we integrated an evolutionary optimization algorithm with this measurement framework to extract semiconductor device parameters from IV measurements of electronic devices, such as diodes. This integration provides an end-to-end measurement and analysis software solution, from automated measurement of current-voltage characteristics, through extracting electronic device parameters, to understanding of semiconductor transport properties (Fig. \ref{fig1}). Finally, we apply this software solution to the characterization of Pt/Cr\textsubscript{2}O\textsubscript{3}:Mg/$\beta$-Ga\textsubscript{2}O\textsubscript{3} heterojunction diode operated at high-temperature. Given the widespread usage of the IV electrical measurement methods and instruments discussed in this work, we shared this software as open-source code on Github, also hoping that it would provide other researchers with a good starting point for more advanced electrical analysis and more tailored feature extraction.

\section{ChatGPT-crafted automation}

\subsection{Controlling the SMU}

To streamline our discussion, we focus on the overarching elements of code development and outcomes, with the comprehensive prompt history detailed in the supplementary materials. Our initial inquiries assessed ChatGPT's ability in crafting Python code and its familiarity with the SCPI (Standard Commands for Programmable Instruments) protocol, needed to control the source measure instrument, as showcased in Fig. \ref{fig2}. 

Despite our prompts containing acronyms and an unintentional typographical error, ChatGPT adeptly navigated these minor obstacles, demonstrating its capability to accurately interpret the context and provide pertinent responses. Moreover, ChatGPT recommended the PyVISA library, which serves as a Python interface to the Virtual Instrument Software Architecture (VISA) API, a standard framework that facilitates communication between software applications and test and measurement equipment.

\begin{figure}[!t]
\centering
\begin{mdframed}[backgroundcolor=gray!20, roundcorner=10pt, innertopmargin=10pt, innerbottommargin=10pt, outerlinewidth=1, linecolor=gray!50]
\textcolor{darkred}{User: Can you run python code?}\\
\textcolor{darkblue}{ChatGPT: Yes, I can run Python code. Feel free to ask for any kind of Python-related task, such as calculations, data analysis, generating graphs, or running simulations. Just let me know what you need, and I'll do my best to assist you!}\\[10pt]
\textcolor{darkred}{User: Are you aware of the SCIPI for instruments, such as Keithley SMUs?}\\
\textcolor{darkblue}{ChatGPT: Yes, I'm aware of SCPI (Standard Commands for Programmable Instruments), which is a standardized language used for controlling programmable test and measurement devices. SCPI is widely used in various instruments, including Keithley Source Measure Units (SMUs). Keithley SMUs, like many other test and measurement devices, often use SCPI commands for automation and control. These commands allow users to set up the instrument, perform measurements, and retrieve data in a standardized way, which is particularly useful in automated test environments. For example, in Python, you might use libraries like PyVISA to communicate with instruments using SCPI commands. PyVISA acts as an interface to the VISA library, which is an industry-standard API for communication with test and measurement equipment. With this setup, you can send SCPI commands to a Keithley SMU to configure it, initiate measurements, and read the results. If you have a specific task or question related to controlling Keithley SMUs or using SCPI commands in Python, feel free to ask!}
\end{mdframed}
\caption{Initial prompts evaluating ChatGPT's comprehension of SCPI (Standard Commands for Programmable Instruments) and its proficiency in generating Python code. The dialogue demonstrates ChatGPT's resilience to a typographical error (SCIPI instead of SCPI), underscoring its robust understanding and ability to accurately respond to queries related to programming and instrument control.}
\label{fig2}
\end{figure}

Subsequently, we tasked ChatGPT with the development of a Python class for managing a Keithley 2400 SMU, detailed in Fig. S1$^\dag$ of the ESI. We provided comprehensive instructions, outlined initial tasks, and suggested internet searches for additional information, leveraging the ability of ChatGPT-4 to augment its responses with web-based data. Additionally, we employed standard terminology relevant to the field, including ``4-wire measurement mode'' and ``compliance level''. Furthermore, we explicitly directed ChatGPT not to execute any code. While ChatGPT-4 is capable of running Python code within its environment, it cannot interface directly with instruments through its conversational interface, and trying to communicate with the SMU would result in VISA communication errors. We could have allowed GPT-4 to remotely control the instrument through OpenAI's API, but this would require human-developed code to handle appropriate parameter input and safety bounds, since it is highly risky to enable code execution from external agents.

Beyond a full comprehension of the task, ChatGPT proposed initial steps for creating a Python class to interface with instruments, presenting a systematic approach for class development that includes initialization, connection, and identification, along with methods for instrument control. It also emphasized the critical requirement of installing PyVISA through \texttt{pip} and the need for a VISA backend, essential for enabling the Python scripts to communicate with the instrument. The complete ChatGPT's answer is shown in Fig. S1 of the ESI$\dag$, and the resulting Python code for this initial step in shown in Fig. \ref{fig3}. It is worth highlighting the understanding of SCPI commands and the requirement to encapsulate them in strings when transmitting them to the instrument through the PyVISA \texttt{write} method. 

\begin{figure}[!t]
\centering
\begin{lstlisting}

class Keithley2400Controller:
    def __init__(self, resource_name):
        self.resource_name = resource_name
        self.instrument = None
        self.current_compliance = 0.01  # 10 mA default

    def connect(self):
        rm = pyvisa.ResourceManager()
        self.instrument = rm.open_resource(self.resource_name)
        self.instrument.write("*RST")  # Reset the instrument
        self.instrument.write("*CLS")  # Clear the status

    def identify(self):
        return self.instrument.query("*IDN?")

    def select_panel(self, panel='FRONT'):
        if panel.upper() == 'FRONT':
            self.instrument.write(":ROUT:TERM FRON")
        elif panel.upper() == 'REAR':
            self.instrument.write(":ROUT:TERM REAR")
        else:
            raise ValueError("Invalid panel option. Choose 'FRONT' or 'REAR'.")

    def set_measurement_mode(self, mode):
        if mode == 2:
            self.instrument.write(":SYST:RSEN OFF")  # 2-wire mode
        elif mode == 4:
            self.instrument.write(":SYST:RSEN ON")  # 4-wire mode
        else:
            raise ValueError("Invalid measurement mode. Choose 2 or 4.")

    # Placeholder for IV sweep method
    # Placeholder for setting sourcing and measurement ranges

    def set_current_compliance(self, compliance):
        self.current_compliance = compliance
        self.instrument.write(f":SENS:CURR:PROT {compliance}")

# Example usage
controller = Keithley2400Controller('GPIB0::24::INSTR')  # Replace with your actual GPIB address
controller.connect()
print(controller.identify())
controller.select_panel('FRONT')
controller.set_measurement_mode(4)
controller.set_current_compliance(0.01)  # 10 mA
\end{lstlisting}
\caption{ChatGPT-crafted Python class (initial architecture, before subsequent refactoring based on prompts) for interacting with a Keithley 2400 SMU, with methods for instrument connection, panel and measurement mode selection, and setting a compliance level. An example usage was also provided.}
\label{fig3}
\end{figure}

We tested this class and example usage, but changing the instrument's GPIB address to the correct one, and we successfully were able to select the front panel, change the measurement mode to a 4-wire mode, and set a current compliance of \SI{10}{\milli\ampere}. Besides the instrument's address --- which we did not pass to ChatGPT and depends on port addressing --- no correction had to be made on the code, and no errors were found during execution, which is notable for a first trial. After this initial test, we informed ChatGPT about the instrument's GPIB address and asked for methods to define current and voltage source and measure ranges, which were successfully implemented and tested, again with no errors. This step is shown in Fig. S2 of the ESI$\dag$. 

The next step was to ask ChatGPT to implement the most challenging part of the control class: implement a current-voltage (IV) sweep. Given the complexity of this task, we provided specific requirements, such as what it should source and measure, buffer set up and initialization, and so on, as shown in Fig. S3 of the ESI$^\dag$. The resulting code missed the command to enable output on the instrument and the previously created methods were not used. After a few iterations to correct these issues and an unexpected output upon testing, we finally decided to provide ChatGPT with an example from the instrument's manual on how to program a linear IV sweep. Details about these interactions can be found in the prompt history.

The IV method was then refactored to take the provided example into account, but accounted only for sourcing current and measuring voltage, and left no room for range options, which was in fact expected, since ChatGPT only reproduced the example. We then instructed ChatGPT to modify this method so we could source and measure both current and voltage, and to not forget to leave compliance, sourcing and measuring ranges, and NPLC as variables, besides making sure to use the methods it had already developed.

After a new round of code refactoring, we had a working version of the IV sweep method. Upon testing, the instrument started the measurements but PyVISA reported a timeout error, due to the fact that the instrument was still executing several measurements for the IV sweep but PyVISA was already expecting an answer. We informed ChatGPT about this error, and it updated the controller class with a timeout parameter that defaults to 25 seconds, thus solving the timeout error. Then, after a successful implementation of the IV sweep method and a few more instructions to handle the resulting data --- with online code execution by ChatGPT to understand and format the data --- and other parameters such as compliance, over-voltage protection and range levels, we finally got a fully working Python class to interact with a Keithley 2400 SMU.

We needed to implement only two code modifications. The first addressed auto-ranging during the IV sweep. We noted that executing the sweep command necessitated a specific instruction for selecting between a fixed range, as pre-configured in the instrument, and auto-range. This aspect was not initially managed correctly by ChatGPT, though further information was not provided for resolution due to time constraints, given it was a minor adjustment. The second modification involved the calculation of sweep points. Specifically, ChatGPT overlooked the necessity of using the absolute value for both the difference between the start and stop levels and the step size, which is crucial for ensuring the linear sweep operates correctly in both ascending and descending modes. Although the inclusion of \texttt{abs(stop - start)} was correct in early iterations of the sweep method, it was inadvertently omitted following subsequent code refactorization. Detailed information about the prompts and the generated Python scripts are accessible via the links in the ESI$^\dag$. 

Finally, although we did not specify anything regarding error handling, it is interesting to note ChatGPT's behavior when dealing with possible errors. For instance, specifying anything different from 2 or 4 for the measurement mode will raise an error since these are the only two allowed parameters, and ChatGPT handled this issue by printing ``Invalid measurement mode. Choose 2 or 4'', as shown in Fig. \ref{fig3}. On the other hand, the \texttt{connect}, \texttt{identify}, and \texttt{iv\_sweep} methods do not handle errors, besides the methods defining range parameters, which is expected since we did not provide any information regarding allowed values.  

\subsection{Graphical user interface (GUI)}

\begin{figure}[!t]
\centering
\begin{mdframed}[backgroundcolor=gray!20, roundcorner=10pt, innertopmargin=10pt, innerbottommargin=10pt, outerlinewidth=1, linecolor=gray!50]
\textcolor{darkred}{User: I'll give you the code for a backend control code named "Keithley2400Controller" stored in a file named "Keithley2400GPT.py". Your task is to build a professional-looking GUI for this code. Here are all things this GUI should accomplish:}
\begin{itemize}
    \setlength\itemsep{0.1em}
    \item \textcolor{darkred}{Connect to the instrument and check connection}
    \item \textcolor{darkred}{A frame for all setup parameters to be filled by the user, except overvoltage, which always defaults to 20 V in the backend code. These parameters are source type, measure type, start level, stop level, step level, compliance, nplc, source delay. For auto range, a toggle that defaults to auto range enabled should be present, but when it's disabled, the user should entry source and measure ranges.}
    \item \textcolor{darkred}{A button to perform the IV sweep based on all configuration parameters. The IV sweep should be executed asynchronously.}
    \item \textcolor{darkred}{A plot area, to the right side of the configuration and setup frame. The data should be displayed only after the IV sweep is completed. }
    \item \textcolor{darkred}{Toggle to enabled y and x-axis log scale. Upon enabling/disabling this toggle, the plot should be dynamically updated.}
    \item \textcolor{darkred}{Button to save the data from the IV sweep. Voltage values in the first column, and current values in the second column.}
\end{itemize}
\textcolor{darkred}{Is it all clear? If so, I'll give you the code to control the instrument.}\\
\textcolor{darkblue}{ChatGPT: Yes, I can help you design a GUI for your backend control code for the Keithley 2400. Please go ahead and share the code for the "Keithley2400Controller". This will help me understand how the backend is structured and how the GUI should interact with it. Once I have the code, I can provide you with a detailed design and even some sample code for the GUI using a Python library like Tkinter or PyQt, depending on your preference.}
\end{mdframed}
\caption{Initial prompt instructing ChatGPT to create a graphical user interface (GUI) for interfacing with the instrument's control class, to assess whether it correctly comprehended its task.}
\label{fig4}
\end{figure}

GUIs are especially useful for quick measurements and for most of materials science researchers that are usually not familiar with scripting programming languages. Thus, we decided to instruct ChatGPT to design and implement a minimal graphical user interface for the instrument control class. For that, we started a new chat and directed ChatGPT to develop a GUI for our instrument control module. We focused on outlining the required functionalities and ensuring ChatGPT comprehended the project scope, as illustrated in Fig. \ref{fig4}.

We copied the control module to the prompt, after which ChatGPT answered with a design for the required application and started generating a code snippet, but mentioned that \textit{``It doesn’t include detailed implementations of event handlers and plotting''}, as recorded in the prompt log. Given that, we stopped the answer and asked ChatGPT to provide a detailed and complete implementation. After this additional instruction, ChatGPT answered with a step-by-step breakdown of the code, as shown in the prompt history available in ESI$^\dag$.

Our previous attempts to develop this GUI were functional but visually unappealing, featuring misplaced buttons and excessive empty spaces. Learning from these experiences, we provided detailed specifications for layout, frame alignment, and button placement, which led to an extensive conversation until the GUI was functional and aesthetically pleasing, as evidenced by the chat log in the supplementary materials. This process resulted in ChatGPT revising the code many times, which inadvertently led to the loss of some essential methods and functionalities. For instance, when corrections were made to the plotting function, updates often overlooked the need for logarithmic values when the logarithmic scale toggle was enabled. Upon identifying code inconsistencies, we decided to make targeted updates only where necessary, rather than revising entire methods. Also, given that we were having some issues with the plot method regarding the axes, we only tilted the x-axis ticks by \SI{45}{\degree} and let \texttt{matplotlib} handle the displays of the ticks. Details of this back-and-forth interaction are recorded in the complete prompt history available through the shared chat links in the ESI$^\dag$.

\begin{figure}[!t]
    \centering
    \includegraphics[width=\columnwidth]{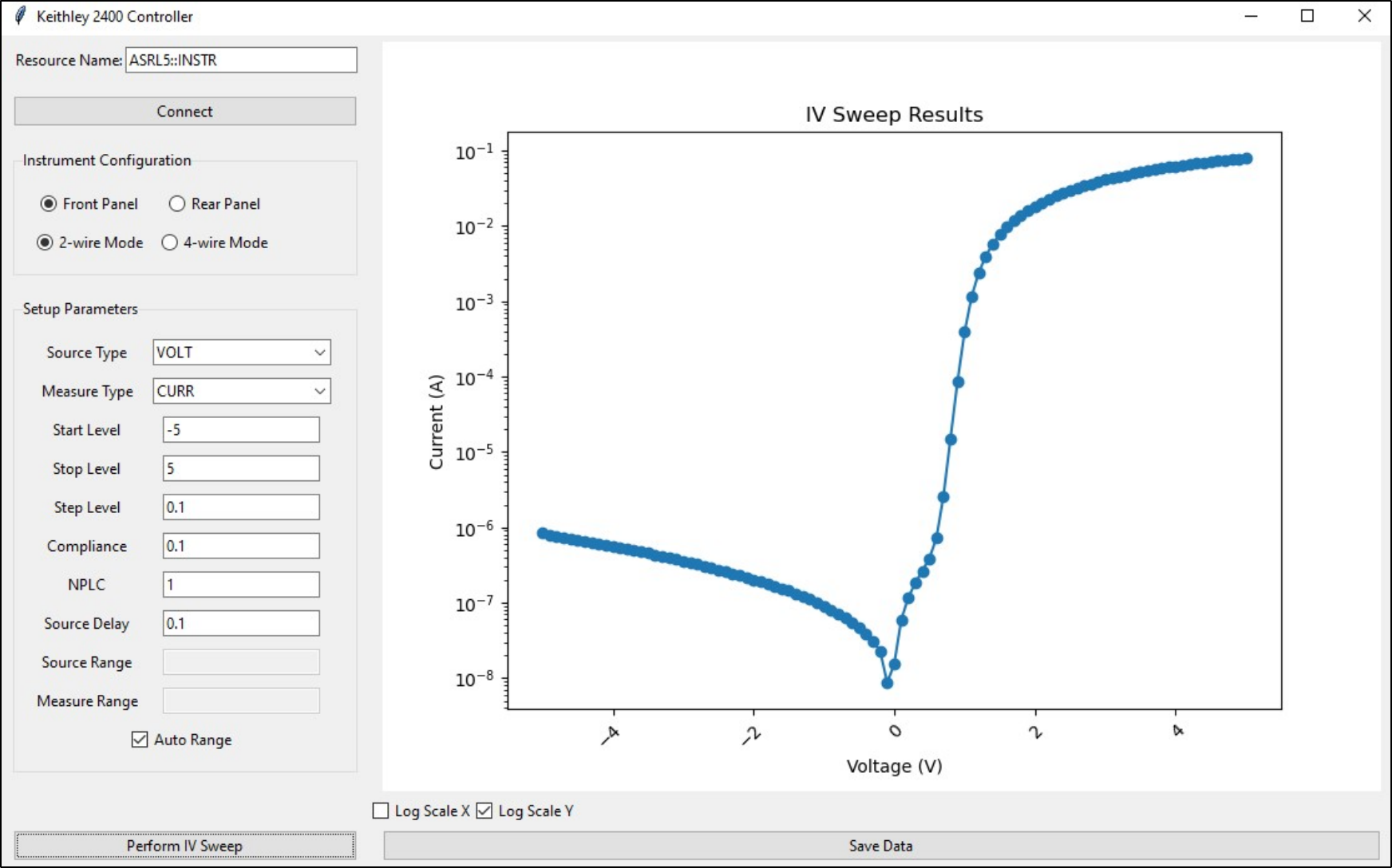}
    \caption{ChatGPT-crafted graphical user interface for a Keithley 2400 controller. The user can select the operation panel, measurement mode, source and measure varibles (voltage and current), sweep parameters (start, stop, step size), compliance, source and measurement ranges, besides NPLC and delay parameters. The data resulting from the IV sweep is plotted and can be saved through a file dialog.}
    \label{fig5}
\end{figure}

Finally, ChatGPT did not handle error exceptions well for the GUI, since the IV sweep and save data button are still enabled even if the instrument is not connected and no data has been collected. Thus, we provided the full GUI code to ChatGPT and asked for appropriate error handling measures for the connect button, save data and perform IV sweep buttons, specifying that these buttons should be disabled if the instrument is not connected, and that the save data button should be disabled if no data were collected. We also required the same for panel and measurement mode toggles. All of these additional features were successfully implemented, as recorded in the prompt history available in the ESI$^\dag$, and we updated the GUI code. The final GUI, shown in Fig. \ref{fig5}, features all necessary control parameters for acquiring current and voltage data from the instrument using linear sweeps, i.e., when the sourced values are equally spaced, including panel selection, measurement mode (2 or 4-wires), range specifications (including auto-range), compliance, measurement speed and source delay. Moreover, the resulting data is displayed on the screen, and the user can save data as a text file. 

\section{Streamlining IV parameter extraction with metaheuristic algorithms}
To automate and speed up the parameter extraction from IV curves, we developed an open-source optimization tool that leverages the self-adaptive differential evolution (DE) algorithm \cite{brest_self-adapting_2006}, and Numba, a high-performance Python compiler that allows easy code parallelization across multiple CPUs\cite{numba2015}.  This tool can be applied to single-objective optimization problems by simply defining an objective function and a search space that are then passed to the model. Although many implementations of the standard differential evolution algorithm are available, such as in \textit{scipy} and \textit{Mathematica}, we have not found any open-source Python implementation of the self-adaptive DE algorithm --- only a \texttt{C++} implementation (pagmo\cite{biscani_parallel_2020}) was found. Extensive discussions and an application of this optimization algorithm to the PV parameter extraction problem can be found elsewhere\cite{brest_self-adapting_2006,febba_effects_2020,febba_effects_2021}.

We briefly describe the features of this tool, but more details can be found in the Github repository listed in the data availability section. We illustrate the usage of the optimization tool to find the minimum of the 2D Ackley function, which possesses many local minima and is widely used to benchmark optimization algorithms, where the goal is to find $(x,y)$ that minimize the Ackley function defined in Eq. (\ref{ackley}):

\begin{equation}
\begin{aligned}
f(x, y) = & -20 \exp\left[-0.2 \sqrt{0.5 (x^2 + y^2)}\right] \\
          & - \exp\left[0.5 (\cos(2\pi x) + \cos(2\pi y))\right] + e + 20
\end{aligned}
\label{ackley}
\end{equation}

Fig. \ref{fig6}a shows a code snippet for the minimization of the 2D Ackley function. As long as the objective function is Numba-compatible and its arguments are \texttt{indv} only, any function can be passed to the \texttt{parallel} method from class \texttt{DE}. The search space is defined by setting lower and upper bounds for each element of the \texttt{indv} array, which contains the parameters of the objective function. 

\begin{figure}[!t]
% Part a)
{\raggedright a) 2D Ackley Function Optimization}\par
\begin{lstlisting}
import DE
from numba import jit
import numpy as np

@jit(nopython=True)
def ackley(indv):
    term1 = -20 * np.exp(-0.2 * np.sqrt(0.5 * (indv[0]**2 + indv[1]**2)))
    term2 = -np.exp(0.5 * (np.cos(2 * np.pi * indv[0]) + np.cos(2 * np.pi * indv[1])))
    value = term1 + term2 + np.e + 20
    return value

lbound = np.array([-10,-10])
ubound = np.array([10,10])

score_list, solution_list = DE.parallel(ackley,runs=1, popsize=100, gmax=1000,lower_bound=lbound, upper_bound=ubound)
\end{lstlisting}

% Part b)
{\raggedright b) Objective Function with External Data}\par
\begin{lstlisting}
import DE
from numba import jit

def create_objective_function(voltage, current):
    """Create an objective function with external data."""
    @jit(nopython=True)
    def objective(indv):
        model = ...
        fitness_scalar = ...
        return fitness_scalar
    return objective

lbound = ...
ubound = ...

voltage, current = DE.load_raw_data("iv.txt", vmin=-2, vmax=2)    
custom_objective = create_objective_function(voltage, current)

score_list, solution_list = DE.parallel(custom_objective,runs=1, popsize=100, gmax=1000,lower_bound=lbound, upper_bound=ubound)
\end{lstlisting}
\caption{a) Code snippet showing the usage of the optimization tool described in this work. In this example, the goal is to mind the minimum of the 2D Ackley function, widely used to benchmark optimization algorithms. b) Definition of an objective function with external data (current and voltage) using a function closure, returning an objective function that depends only on the optimization individuals (\texttt{indv}), to keep the differential evolution nomenclature.}
\label{fig6}
\end{figure}

The term \textit{individual} (\texttt{indv} in the code snippets) refers to the usual nomenclature in the differential evolution literature, but it is nothing more than a possible solution for the optimization problem --- each component of an individual is a parameter that will be optimized. In the example of Fig. 6a, \texttt{indv[0]} represents the \textit{x} variable, while \texttt{indv[1]} is the \textit{y} variable, so that we have the 2D Ackley function, and the minimum and maximum allowed values of each dimension are defined by elements 0 and 1 of the lower and upper bound arrays. If the objective function needs more arguments or external data, such as voltage and current, we recommend using a function enclosure, where a higher-order function that can take additional arguments encapsulates the objective function, which serves as a closure, as shown in Fig. \ref{fig6}b. Moreover, our optimization tool leverages the easy parallelization enabled by Numba: the parameter \texttt{runs} controls the number of parallel optimization processes, each starting with a different \texttt{seed} parameter, so that each \textit{evolution} of parameters starts at different points, to enhance the likelihood of finding the global optimum. 

We measured the optimization performance on the Ackley function on a laptop equipped with a 11th Gen Intel Core i7-11850H (2.50GHz) and 16 GB of RAM, running a 64-bit Windows 10 OS: it took on average 3.9 seconds to run 10 parallel optimization routines (across 100 code executions), each with a population size of 100 individuals evolving until a maximum number of 10,000 generations is reached. The results available in the Git repository (\texttt{ackley\_results.py}) demonstrate that the algorithm successfully found the minimum of the 2D Ackley function, which is 0 at $(x,y) = (0,0)$, by always finding x between \num{-3.6e-16} and \num{3.6e-16}, and y values between \num{-3.6e-16} and \num{3.6e-16}, with the minimum of the Ackley function always converging to \num{4.4e-16}.

\section{Application: modeling Pt/Cr\textsubscript{2}O\textsubscript{3}:Mg/$\beta$-Ga\textsubscript{2}O\textsubscript{3} heterojunction diodes}

\subsection{Device fabrication and IV measurements}

To demonstrate the developed applications for device I-V characteristics parameter extraction and analysis, a Pt/Cr\textsubscript{2}O\textsubscript{3}:Mg/$\beta$-Ga\textsubscript{2}O\textsubscript{3} vertical  heterojunction diode was fabricated using an \SI{7}{\micro\meter} lightly Si-doped (\SI{8e-15}{\per\cubic\cm}) n-type $\beta$-Ga\textsubscript{2}O\textsubscript{3} drift layer grown on a conductive bulk (001) $\beta$-Ga\textsubscript{2}O\textsubscript{3} substrate (NCT). The schematic of the device cross-section is shown in inset of Fig. \ref{fitting}. Before device fabrication, photoresist was removed from the as-delivered substrates via an organic wash followed by a piranha rinse. Large area ohmic contact of Ti/Au (\SI{5}{\nano\meter}/\SI{100}{\nano\meter})\cite{callahan_ultrathin_2023} was deposited on the backside of the substrate via e-beam evaporation, followed by a rapid thermal annealing (RTA) in N\textsubscript{2} ambient at \SI{550}{\celsius} for \SI{1}{\min}. Next, the Cr\textsubscript{2}O\textsubscript{3}:Mg layer was grown by pulsed laser deposition(PLD), a ceramic target of 8 at.\% Mg-doped Cr\textsubscript{2}O\textsubscript{3} was ablated using a pulsed KrF excimer laser at a frequency of 10Hz and energy of 300mJ. Growth was performed at a substrate temperature setpoint of \SI{600}{\celsius} and O\textsubscript{2} partial pressure of \SI{3e-4}{Torr}. Details of the electrical properties of the Cr\textsubscript{2}O\textsubscript{3}:Mg layer have been described elsewhere\cite{callahan_reliable_2024}. To make the top contact shadow masks were used to define an array of 1 mm diameter pads and e-beam evaporation is used to deposit a \SI{30}{\nano\meter} Pt layer on the Cr\textsubscript{2}O\textsubscript{3}:Mg. 

Current-voltage data from Pt/Cr\textsubscript{2}O\textsubscript{3}:Mg/$\beta$-Ga\textsubscript{2}O\textsubscript{3} heterojunction diodes were obtained in a high temperature and highly-automated probe station capable of controlling a reducing atmosphere for temperature-dependent performance and time-dependent reliability electrical device measurements. All IV data were obtained with the ChatGPT-developed control module for the Keithley 2400 SMU. In this application, we sourced voltage between \SIrange{-5}{10}{\volt} and measured the resulting current in a 2-wire measurement mode, under normal measurement speed (NPLC = 1). Futhermore, we set current compliance to \SI{200}{\milli\ampere} and used auto-range for both source and measure functions. A single heterojunction diode was measured at various temperatures without H\textsubscript{2} reducing conditions, and at \SI{500}{\celsius} under different gas flow conditions. Two Alicat mass flow controllers (MFCs) were used to control the flow of N\textsubscript{2} and forming gas (2 \% H\textsubscript{2}), which were directed into the test chamber from two separate gas lines that mixed before the chamber feedthrough. The total gas flow was always 100 sccm, with the forming gas flow set at 0 sccm, 2 sccm, and 5 sccm. JV curves were recorded at 1-minute intervals without any initial device equilibration time and were subsequently analyzed using a parameter extraction procedure, as detailed in the following section.

\subsection{Parameter extraction methods}
We present here an application of the developed control code and streamlined optimization tool: the modeling of Pt/Cr\textsubscript{2}O\textsubscript{3}:Mg/$\beta$-Ga\textsubscript{2}O\textsubscript{3} heterojunction devices. The transport properties in these devices can be studied through diode parameters extracted from IV measurements, usually through the single-diode model, defined by Eq. (\ref{sdm}) with a shunt term:

\begin{equation}
    I = I_0\left[\exp{\left(\frac{V-IR_s}{nV_T}\right)} - 1\right] + \frac{V-IR_s}{R_{sh}},
    \label{sdm}
\end{equation}
where I\textsubscript{0}, n, R\textsubscript{s} and R\textsubscript{sh} represent the diode reverse saturation current, ideality factor, series and shunt resistance, respectively, and $V_T=k_{\beta}T/q$ is the thermal voltage. Hence, from each set of IV measurements at a given temperature, the goal is to extract the four diode parameters from Eq. (\ref{sdm}). This equation can be rearranged to express current as a function of voltage and single-diode parameters only, through the Lambert W function\cite{jung_schottky_2009,calasan_root_2020,aazou_lambertw_2022}, as defined by Eq. (\ref{lambert}):

\begin{align}
    I &= \frac{nV_T}{R_s} W_0 \left\{ \frac{R_s I_0}{nV_T \left( 1 + \frac{R_s}{R_{sh}} \right)} \left[ \exp \left( \frac{R_s I_0 + V}{nV_T \left( 1 + \frac{R_s}{R_{sh}} \right)} \right) \right] \right\} \notag \\
    &\quad - \frac{I_0 - \frac{V}{R_{sh}}}{1 + \frac{R_s}{R_{sh}}},
    \label{lambert}
\end{align}
where W\textsubscript{0} is the main branch of the Lambert W function. Additionally, we considered a voltage offset $V = V + V_{offset}$ to account for additional voltage inaccuracies from the measurement setup. Thus, we can simulate IV curves based on single-diode parameters and voltage, and then compare how close simulated curves are to the experimental ones by defining an objective metric, which will be minimized by the optimization algorithm. In this case, we used the Root Mean Squared Error, but taking the relative deviation to account for the orders of magnitude variation in the current, as defined by Eq.\ref{rmse}:

\begin{equation}
    RMSE = \sqrt{\frac{1}{N}\sum_{i=1}^N\left(\frac{I_{i,exp} - I_{i,sim}}{I_{i,exp}}\right)^2},
    \label{rmse}
\end{equation}
where I\textsubscript{i,sim} is the i-th simulated current point obtained from Eq. (\ref{lambert}), I\textsubscript{i,exp} are the experimentally measured current points and N is the number of experimental IV points.

Therefore, the goal of the optimization algorithm is to find single-diode parameters that minimize the RMSE. However, since we used Numba to compile the DE algorithm, we can only use Numba-compatible functions when defining an objective function. Since off-the-shelf implementations such as \texttt{scipy.special.lambertw} are not Numba-compatible, we approximated the main branch of the Lambert W function, $w=W_0(x)$, through the Newton-Rhapson method\cite{loczi_guaranteed-_2022} as given by Eq. (\ref{newton}):
\begin{equation}
    w_{i+1}(x) = w_i(x) - \frac{w_i(x)e^{w_i(x)}-x}{e^{w_i(x)}+w_i(x)e^{w_i(x)}},
    \label{newton}
\end{equation}
where a rough initial guess is given by $w_i(x)=ln(x+1)$. In this case, we found that 10 iterations are sufficient to approximate W\textsubscript{0} within an absolute error lower than \num{1e-10} in the interval $[\num{e-10},\num{e6}]$. An implementation to extract single-diode parameters from IV data using the applications reported here is provided in our remote repository$^\dag$, as well as the measured IV data and the results of the Newthon-Raphson approximation of the Lambert W function.

\subsection{Results}

The developed tool described in this work makes it straightforward to extract diode parameters from IV curves by just defining search bounds and Eq. (\ref{rmse}) as a custom objective function through a function closure, as shown in Fig. \ref{fig6}b, and in the parameter extraction script available in our Git repository. Thus, IV data obtained with the ChatGPT-crafted control module is piped to the parameter extraction tool, and in a few seconds all relevant diode parameters are available.

\begin{figure}[!t]
    \centering
    \includegraphics[width=\columnwidth]{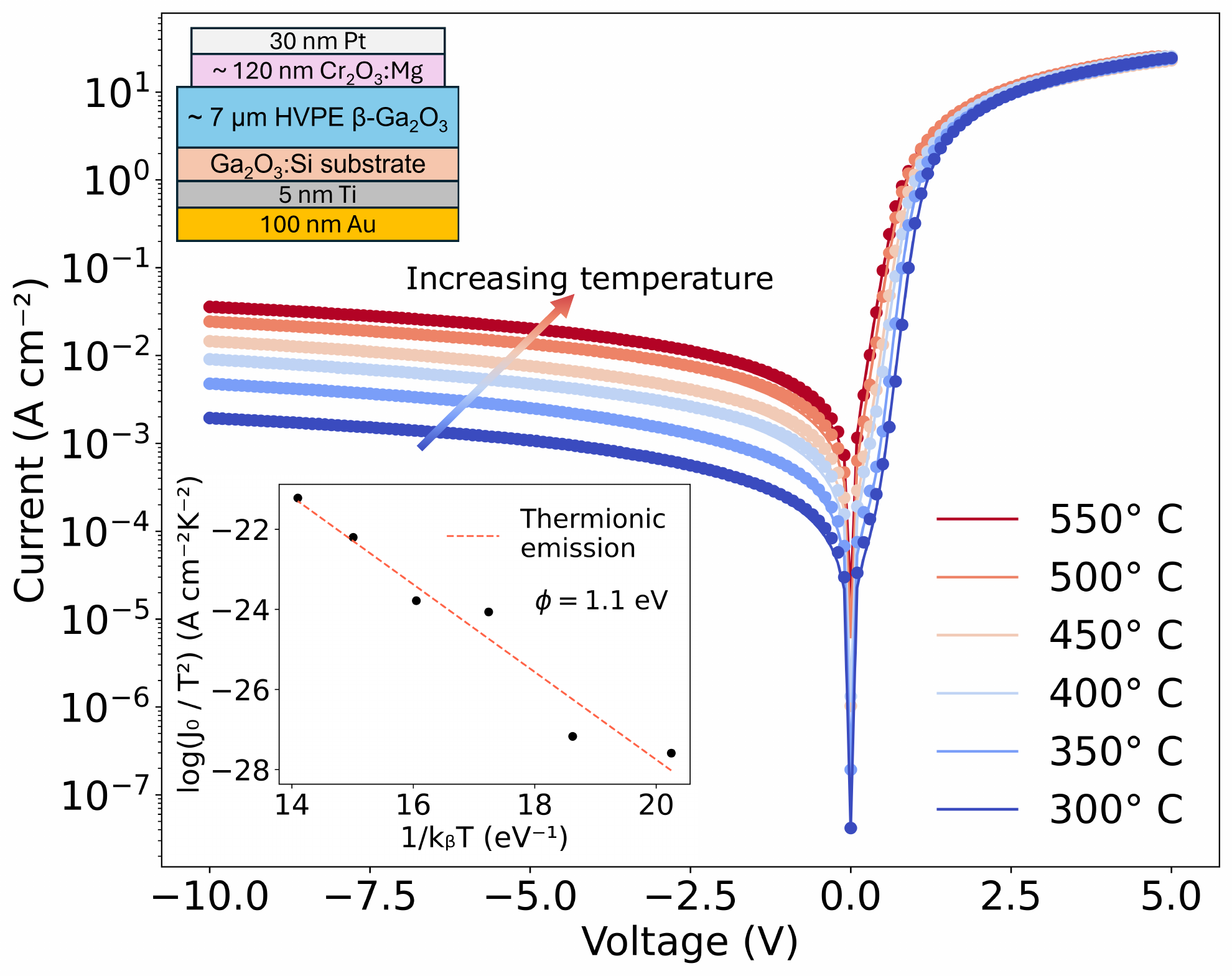}
    \caption{Architecture of a Pt/Cr\textsubscript{2}O\textsubscript{3}:Mg/$\beta$-Ga\textsubscript{2}O-based heterojunction diode and corresponding current-voltage data measured at several temperatures. Points represent experimental data and the solid lines represent the fits with the best set of single diode parameters that models the data. MAPEs for ranged between 4.7\% and 7.8\%. The inset shows that J\textsubscript{0} can be modelled by the thermionic emission model, yielding a barrier height of \SI{1.1}{\electronvolt}.}
    \label{fitting}
\end{figure}

For this application case, Fig. \ref{fitting} shows the IV data measured at several temperatures and the resulting fitting from \SIrange{-10}{5}{\volt} for a Pt/Cr\textsubscript{2}O\textsubscript{3}:Mg/$\beta$-Ga\textsubscript{2}O\textsubscript{3}-based heterojunction diode in the \SIrange[range-phrase=--]{300}{550}{\celsius} temperature range under trace H\textsubscript{2} concentration, using the ChatGPT-crafted control module for Keithley 2400 SMUs. Mean absolute percentage errors (MAPE) ranged between 4.7\% and 7.8\%, demonstrating that the parameter extraction tool was able to successfully extract diode parameters from the IV curves.

As shown in the inset of Fig. \ref{fitting}, the extracted reverse saturation current J\textsubscript{0} can be modelled by the thermionic emission model, defined by Eq. (\ref{thermionic}):
\begin{equation}
    J_0=A^*T^2e^{-\phi/k_{\beta}T},
    \label{thermionic}
\end{equation}
where A\textsuperscript{*} is the effective Richardson constant and $\phi$ is the barrier height, found to be \SI{1.1}{\electronvolt} for this device, which is similar to what has been reported for $\beta$-Ga\textsubscript{2}O\textsubscript{3} Schottky devices \cite{yao_electrical_2017,heinselman_performance_2021} and lower than NiO/$\beta$-Ga\textsubscript{2}O\textsubscript{3} heterojunction devices \cite{sohel_gallium_2023}. Poole-Frenkel emission was found to account for the electric field dependence of J\textsubscript{0} in Pt/Cr\textsubscript{2}O\textsubscript{3}:Mg/$\beta$-Ga\textsubscript{2}O devices, with unintentional Mg-rich layer at the heterojunction interface \cite{callahan_reliable_2024}. Since both thermionic emission and trap-assisted transport can explain the temperature- and field-dependence of current, it appears that both transport mechanisms are present, i.e., activation from the conduction band over the barrier at the junction, and activated transport through trap states in the Mg-rich interfacial layer.

\begin{figure}[!t]
    \centering
    \includegraphics[width=\columnwidth]{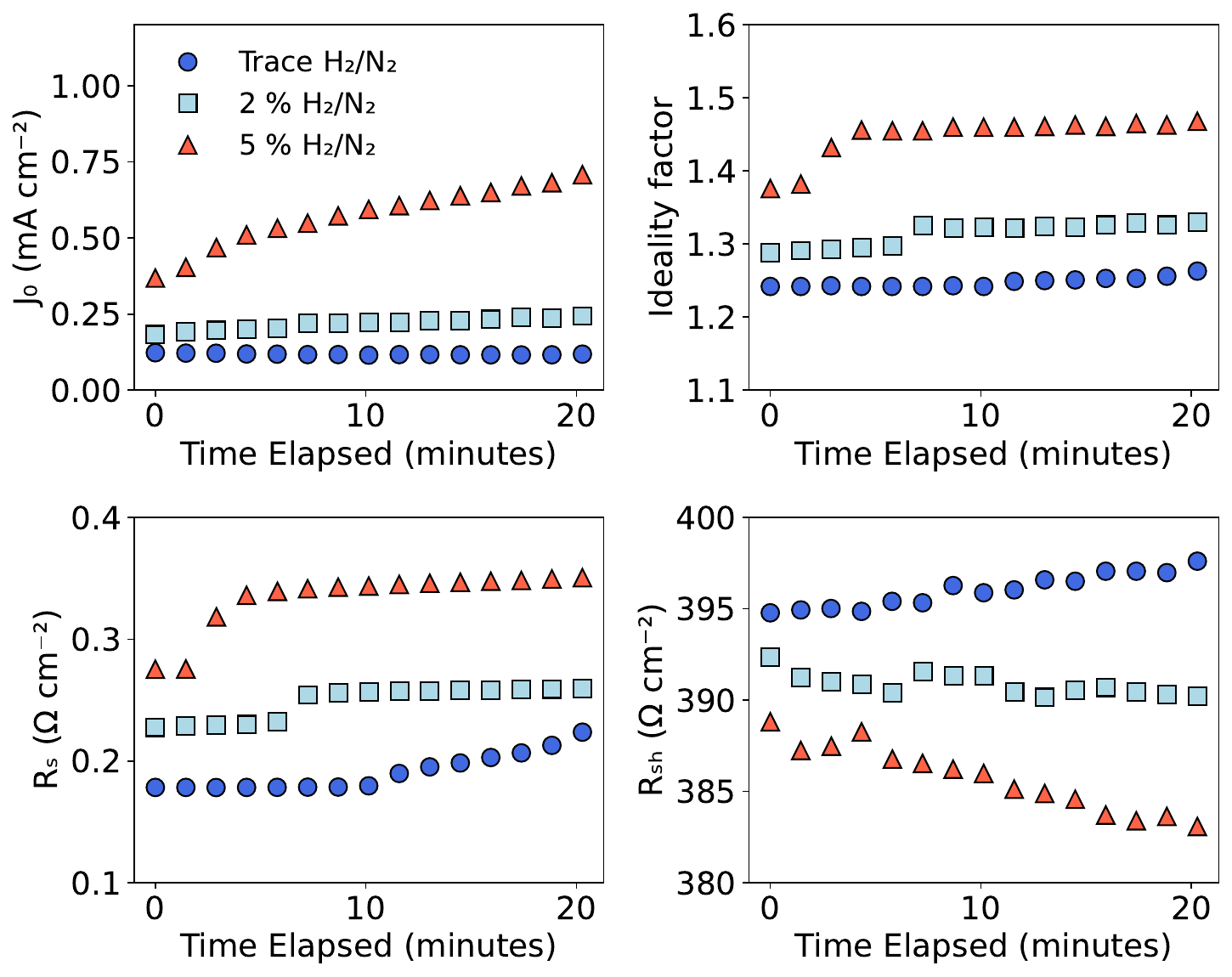}
    \caption{Single-diode parameters as a function of time and forming gas percentage in the N\textsubscript{2} + H\textsubscript{2}/N\textsubscript{2} mixture for the tested Cr\textsubscript{2}O\textsubscript{3}/$\beta$-Ga\textsubscript{2}O-based heterojunction diode. Device degradation was observed upon exposure to N\textsubscript{2}H\textsubscript{2} while enduring a temperature soak at \SI{500}{\celsius}.}
    \label{parameters_time}
\end{figure}

Moreover, we investigated the device's response to H\textsubscript{2} atmosphere conditions, by extracting the diode parameters at several H\textsubscript{2} concentrations over time, as shown in Fig. \ref{parameters_time}. Fig. \ref{parameters_time}a shows the reverse saturation current extracted from IV curves measured at different H\textsubscript{2} concentrations as a function of time. The increase in $J_0$ with H\textsubscript{2} can be attributed to a decrease of the potential barrier for thermionic emission\cite{lee_sensing_2010}. On the other hand, the steady rise in J\textsubscript{0} over time stems from the device's inability to attain an equilibrium state within the observed time frame for the applied hydrogen concentrations, likely due to the large volume of the measurement chamber. Moreover, a departure from ideal diode behavior as H\textsubscript{2} increases (Fig. \ref{parameters_time}b), with a corresponding increase in lumped series resistance (Fig. \ref{parameters_time}c), and a decrease in shunt resistance (Fig. \ref{parameters_time}d) show that the device was degrading upon exposure to H\textsubscript{2} in harsh environmental conditions.

We measured the performance of our software package for this parameter extraction problem with the same laptop used to minimize the Ackley function, and it took on average 58.7 seconds to run 10 parallel optimization routines with external data reading (single IV curve), each with a population size of 100 individuals evolving until a maximum number of 10,000 generations is reached. We note that the computing time will change according to the calculations required to compute the objective function. Here, Eq. (\ref{newton}) is used to approximate the Lambert W function for every voltage point for each individual, which considerably increases the computational burden --- as a comparison, it took less than 10 seconds to minimize the Ackley function for the same optimization parameters. Still, 10 parallel evolutionary optimization processes can be completed in under \SI{1}{\min}.

Finally, the amount of data can quickly increase for reliability analysis, when IV curves are measured over extended periods of time under several gas concentrations. For these cases, a manual parameter extraction based on curve fitting becomes impractical. Hence, having a tool that can quickly extract diode parameters with minimal user effort can considerably help speed up the characterization process,and take the human out of the loop for fully autonomous accelerated degradation studies of electronic devices. Finally, we mention that complete implementation examples and the code used to generate the figures shown in this work can be found in the ESI$^\dag$.

\section{Conclusions}
In this study, we automated a Keithley 2400 Source Measure Unit (SMU) through an interactive dialogue-with ChatGPT-4, that required just a few hours of time and no human-developed code besides few minor corrections. This software development method significantly streamlined the instrumental setup and testing phases, allowing researchers to focus on getting and analyzing materials science and device engineering results, rather than grappling with programming tasks. Furthermore, the development of a graphical user interface (GUI) as part of this automation process, which is notorious for having a tedious software development process, enhanced the user experience with the measurement instrument, making it more intuitive and accessible than the programming language, which is especially beneficial for materials science researchers with little scripting practice. 

Furthermore,  we developed a Python-based implementation of the self-adaptive differential evolution algorithm for parameter extraction analysis of IV electrical measurement results, considering the widespread use of this data analysis approach. This analysis software implementation is enhanced by Numba, a just-in-time compiler that transforms Python code into machine code, significantly accelerating the parameter extraction process from IV curves. By showing a real-world application of this software platform to parameter extraction from IV measurements of Pt/Cr\textsubscript{2}O\textsubscript{3}:Mg/$\beta$-Ga\textsubscript{2}O devices, we demonstrate that AI-driven laboratory automation can quickly yield insights into semiconductor device physics. 

The AI-engineered control module, along with the user-friendly graphical user interface (GUI) and the parameter extraction algorithm described in this paper, are made open-source through Github, so that the whole community can benefit from and contribute to the further development of these tools. Looking ahead, this LMM-based software development method holds the potential to revolutionize research automation by enabling the control of a diverse spectrum of laboratory instruments. Potential future applications include but are not limited to multimeters, temperature controllers, power supplies, mass flow controllers, programmable logic controllers, or any other equipment that features standard communication and command protocols such as Modbus, OPC UA, SCPI. The resulting increased laboratory automation would pave the way for a more interconnected and efficient research environment through the use of Large Language Models.

\section*{Data availability}
Source code for the ChatGPT-crafted control module and GUI is available at:
\href{https://github.com/NREL/Keithley_GPT}{https://github.com/NREL/Keithley\_GPT}. The Differential Evolution algorithm source code and Jupyter notebooks used to analyze the results and generate the figures in this paper are available at: \href{https://github.com/NREL/DE}{https://github.com/NREL/DE}.

\section*{Author contributions}
D. Febba and Andriy Zakutayev conceptualized the system described in this work. D. Febba drafted the manuscript with input from all the co-authors, and operated the characterization system. K. Egbo fabricated the heterojunction devices. D. Febba and W. Callahan tested the probe station atmosphere control and scripting routines. ChatGPT (GPT-4 version) was used to craft Python code and to refine the text. The final version of this manuscript was approved by all authors.

\section*{Conflicts of interest}
There are no conflicts to declare.

\section*{Acknowledgements}
This work was authored by the National Renewable Energy Laboratory, operated by Alliance for Sustainable Energy, LLC, for the U.S. Department of Energy (DOE) under Contract No. DE-AC36-08GO28308. Funding is provided by the Office of Energy Efficiency and Renewable Energy (EERE) Advanced Materials \& Manufacturing Technologies Office (AMMTO) (IV parameter extraction and semiconductor device measurements); and by a laboratory directed research and development (LDRD) program at NREL (AI-enhanced development of instrument control and graphical user interface) The views expressed in the article do not necessarily represent the views of the DOE or the U.S. Government.

\bibliography{references}
\clearpage
\onecolumngrid
\section*{Electronic Supplementary Material}
%%%%%%%%%%%%%%%%%%%%%%%%%%%%%%%%%%%%%%%%%%%%%%%%%%%%%%%%%%%%%%%%%%%%%%
% writeLaTeX Example: Academic Paper Template
%
% Source: http://www.writelatex.com
% 
% Feel free to distribute this example, but please keep the referral
% to writelatex.com
% 
%%%%%%%%%%%%%%%%%%%%%%%%%%%%%%%%%%%%%%%%%%%%%%%%%%%%%%%%%%%%%%%%%%%%%%
% How to use writeLaTeX: 
%
% You edit the source code here on the left, and the preview on the
% right shows you the result within a few seconds.
%
% Bookmark this page and share the URL with your co-authors. They can
% edit at the same time!
%
% You can upload figures, bibliographies, custom classes and
% styles using the files menu.
%
% If you're new to LaTeX, the wikibook is a great place to start:
% http://en.wikibooks.org/wiki/LaTeX
%
%%%%%%%%%%%%%%%%%%%%%%%%%%%%%%%%%%%%%%%%%%%%%%%%%%%%%%%%%%%%%%%%%%%%%%
%\usepackage{epstopdf}%This line makes .eps figures into .pdf - please comment out if not required.
\renewcommand{\thefigure}{ES-\arabic{figure}}
\setcounter{figure}{0} 

\definecolor{cream}{RGB}{222,217,201}

% Define custom colors
\definecolor{codebg}{rgb}{0.95, 0.95, 0.95}
\definecolor{codeblue}{rgb}{0.2, 0.2, 0.6}
\definecolor{codegreen}{rgb}{0, 0.6, 0}
\definecolor{codered}{rgb}{0.6, 0, 0}
 
\lstdefinestyle{mystyle}{
    backgroundcolor=\color{codebg},   % Background color
    basicstyle=\ttfamily\footnotesize, % Basic font style
    keywordstyle=\color{codeblue}\bfseries, % Keyword style
    commentstyle=\color{codegreen}\itshape, % Comment style
    stringstyle=\color{codered},      % String style
    numberstyle=\tiny\color{gray},    % Line number style
    stepnumber=1,                     % Step number for line numbers
    numbersep=5pt,                    % Distance between line numbers and code
    showspaces=false,                 % Do not show spaces
    showstringspaces=false,           % Do not show string spaces
    showtabs=false,                   % Do not show tabs
    frame=single,                     % Single line frame
    rulecolor=\color{black},          % Frame color
    framesep=5pt,                     % Frame separation
    framerule=0.5pt,                  % Frame rule width
    breaklines=true,                  % Enable line breaking
    breakatwhitespace=true,           % Break at whitespace
    tabsize=4,                        % Tab size
    captionpos=b,                     % Caption position (bottom)
    keepspaces=true,                  % Keep spaces
    columns=flexible,                 % Flexible column width
}

\lstset{style=mystyle}

\title{Electronic Supplementary Information}
\maketitle

\lstset{language=Python}

We provide here relevant dialogue between the user and ChatGPT. For full prompt history and code, we refer the reader to the following links:
\begin{itemize}
    \item Python control class: https://chat.openai.com/share/1a9afe99-86ac-499e-94a7-7d0a1a84aa23
    \item GUI: https://chat.openai.com/share/e02fff82-edea-45f3-964d-73ef2012f8ee
    \item GUI (error handling): https://chat.openai.com/share/057d1f7f-61a6-4ce0-9371-8f5b5267625c
    \item ChatGPT-crafted control code and GUI: https://github.com/NREL/Keithley\_GPT
    \item Differential Evolution tool: https://github.com/NREL/DE
\end{itemize}

\begin{figure*}[!h]
\begin{mdframed}[backgroundcolor=gray!20, roundcorner=10pt, innertopmargin=10pt, innerbottommargin=10pt, outerlinewidth=1, linecolor=gray!50]
\textcolor{darkred}{User: Great. I have a Keithley 2400 SMU, that can be controlled through SCPI commands. Using PyVisa, build a python class for controlling this instrument. We should be able to:}
\begin{itemize}
    \setlength{\itemsep}{0pt}
    \item[\textcolor{darkred}{1.}] \textcolor{darkred}{Choose whether to use the front or rear panel.}
    \item[\textcolor{darkred}{2.}] \textcolor{darkred}{Use 2 or 4-wires measurement mode.}
    \item[\textcolor{darkred}{3.}] \textcolor{darkred}{Perform an IV sweep, sourcing voltage and measuring current, with user-defined range for the sweep, and also user-defined measurement and sourcing range (each range has a different accuracy).}
    \item[\textcolor{darkred}{4.}] \textcolor{darkred}{Define a current compliance of 10 mA (default), but also make this a variable.}
\end{itemize}
\textcolor{darkred}{Important to note that this script should be for a Keithley 2400. Browse the internet if needed to find specific commands for this SMU.\\
Do not try to run anything, just report the script back and I'll run it in the instrument and provide feedback, so we can adjust what's wrong. Thus, let's start from basic and initial configuration, such as connecting and sending commands to identify the connection address and the instrument, and then proceed through panel selection and so on. If something is not clear, ask for clarification.}\\
\textcolor{darkblue}{ChatGPT: Certainly! To create a Python class for controlling the Keithley 2400 SMU using PyVISA and SCPI commands, we'll structure the class to include methods for the tasks you've described. I'll begin with a basic structure including initialization, connection, and identification methods, and then add methods for selecting the front or rear panel, setting measurement mode, performing an IV sweep, and defining current compliance. Before running this script, make sure you have PyVISA installed in your Python environment. You can install it using pip: \texttt{pip install pyvisa}\\
Also, you'll need the VISA backend, like NI-VISA or Keysight VISA, installed on your computer to interface with the instrument.}\\
Python code shown in Fig. 2 of the manuscript.\\
\textcolor{darkblue}{ChatGPT: This script provides a basic structure for your Keithley 2400 controller class. You'll need to replace 'GPIB0::24::INSTR' with the actual GPIB address or connection string of your Keithley 2400. Next steps would include adding methods for IV sweep with user-defined parameters. Let me know if this initial setup works for you, and then we can proceed with the implementation of the IV sweep functionality.}
\end{mdframed}
\caption{Dialogue between the user and ChatGPT outlining the development of a Python class for the Keithley 2400 SMU control via PyVISA and SCPI commands, including the ability to select panel types, measurement modes, perform IV sweeps, and set current compliance, followed by ChatGPT's affirmative response detailing the class structure and preliminary steps, emphasizing the necessity of PyVISA installation.}
\label{s1}
\end{figure*}

\newpage

\begin{figure*}[!h]
\begin{mdframed}[backgroundcolor=gray!20, roundcorner=10pt, innertopmargin=10pt, innerbottommargin=10pt, outerlinewidth=1, linecolor=gray!50]
\textcolor{darkred}{User: Great, it works. Now,implement methods to define current and voltage ranges, either for sourcing and measuring cases.}\\
\textcolor{darkblue}{ChatGPT: Fantastic! Let's move on to implementing methods for defining current and voltage ranges in the Keithley2400Controller class. We'll add two sets of methods: one for setting the current range and another for setting the voltage range. Each set will include methods for both sourcing and measuring. The SCPI commands for setting the current and voltage ranges are typically structured as follows:}
\begin{itemize}
    \item \textcolor{darkblue}{For setting the source current range: :SOUR:CURR:RANG $<value>$}
    \item \textcolor{darkblue}{For setting the source voltage range: :SOUR:VOLT:RANG $<value>$}
    \item \textcolor{darkblue}{For setting the current measurement range: :SENS:CURR:RANG $<value>$}
    \item \textcolor{darkblue}{For setting the voltage measurement range: :SENS:VOLT:RANG $<value>$}
\end{itemize}
\textcolor{darkblue}{Here is how these methods can be implemented in the class:}\\
\begin{lstlisting}
class Keithley2400Controller:
    def __init__(self, resource_name='ASRL5::INSTR'):
        # ... existing initialization code ...
    def connect(self):
        # ... existing connection code ...
    def identify(self):
        # ... existing identification code ...
    def select_panel(self, panel='FRONT'):
        # ... existing code ...
    def set_measurement_mode(self, mode):
        # ... existing code ...
    def set_current_compliance(self, compliance):
        # ... existing code ...
    def set_nplc(self, nplc, measurement_type='CURR'):
        # ... existing code ...
    def set_source_current_range(self, range_value):
        self.instrument.write(f":SOUR:CURR:RANG {range_value}")
    def set_source_voltage_range(self, range_value):
        self.instrument.write(f":SOUR:VOLT:RANG {range_value}")
    def set_measure_current_range(self, range_value):
        self.instrument.write(f":SENS:CURR:RANG {range_value}")
    def set_measure_voltage_range(self, range_value):
        self.instrument.write(f":SENS:VOLT:RANG {range_value}")
# Example usage
controller = Keithley2400Controller()
controller.connect()
controller.set_source_current_range(0.1)  # Set source current range to 0.1 A
controller.set_source_voltage_range(20)   # Set source voltage range to 20 V
controller.set_measure_current_range(0.1) # Set measure current range to 0.1 A
controller.set_measure_voltage_range(20)  # Set measure voltage range to 20 V
\end{lstlisting}
\textcolor{darkblue}{ChatGPT: Each of these methods takes a single parameter, \texttt{range\char`_value}, which is the desired range for sourcing or measuring current or voltage. The units should be in amperes for current and volts for voltage. Test these methods with your Keithley 2400 SMU to ensure they work as expected. If you have any more functionalities to add or if any adjustments are needed, please let me know!}
\end{mdframed}
\caption{User-ChatGPT interaction to implement additional methods to set measure and source ranges for voltage and current, and resulting code.}
\label{s2}
\end{figure*}

\begin{figure*}[h!]
\begin{mdframed}[backgroundcolor=gray!20, roundcorner=10pt, innertopmargin=10pt, innerbottommargin=10pt, outerlinewidth=1, linecolor=gray!50]
\textcolor{darkred}{User: Great, it works. Now, implement a method to perform an IV sweep, either by sourcing voltage or current, and either by measuring voltage or current. For example, this method should be capable to source voltage and measure current across a user-defined voltage interval, and the same for sourcing current and measuring voltage. Also, the step size of the sweep should be user-defined. For this task, it may be necessary to configure and initialize the instrument buffer with specific size. If you're not clear about this task, ask for clarification.}\\
\textcolor{darkblue}{ChatGPT: I'm glad to hear that the previous implementations are working well. For the IV sweep functionality, we'll create a method that can handle both voltage sourcing with current measurement and current sourcing with voltage measurement. The method will allow users to define the start and end points of the sweep, the step size, and the type of sweep (voltage or current sourcing). In SCPI, setting up an IV sweep typically involves configuring the source function, setting the start and stop levels, the step size, and the measurement parameters. We also need to set up the buffer to store the measurement results. The buffer size will depend on the number of points in the sweep, which we can calculate from the start, stop, and step values.\\
Here's how you can implement the IV sweep method:}
\textcolor{darkblue}{Here is how these methods can be implemented in the class:}
\begin{lstlisting}
class Keithley2400Controller:
    # ... existing methods ...
    def iv_sweep(self, start, stop, step, source_type='VOLT', measure_type='CURR'):
        # Calculate number of points in the sweep
        num_points = int(abs(stop - start) / step) + 1
        # Set up source and measurement
        if source_type.upper() == 'VOLT' and measure_type.upper() == 'CURR':
            self.instrument.write(":SOUR:FUNC VOLT")
            self.instrument.write(f":SOUR:VOLT:START {start}")
            self.instrument.write(f":SOUR:VOLT:STOP {stop}")
            self.instrument.write(f":SOUR:VOLT:STEP {step}")
            self.instrument.write(":SOUR:SWE:RANG AUTO")
            self.instrument.write(":SOUR:SWE:SPAC LIN")
            self.instrument.write(f":SOUR:SWE:POIN {num_points}")
            self.instrument.write(":SOUR:SWE:DIR UP")
            self.instrument.write(":SENS:FUNC 'CURR:DC'")
        elif source_type.upper() == 'CURR' and measure_type.upper() == 'VOLT':
            self.instrument.write(":SOUR:FUNC CURR")
            self.instrument.write(f":SOUR:CURR:START {start}")
            self.instrument.write(f":SOUR:CURR:STOP {stop}")
            self.instrument.write(f":SOUR:CURR:STEP {step}")
            self.instrument.write(":SOUR:SWE:RANG AUTO")
            self.instrument.write(":SOUR:SWE:SPAC LIN")
            self.instrument.write(f":SOUR:SWE:POIN {num_points}")
            self.instrument.write(":SOUR:SWE:DIR UP")
            self.instrument.write(":SENS:FUNC 'VOLT:DC'")
        else:
            raise ValueError("Invalid source/measure type combination. Choose 'VOLT'/'CURR' or 'CURR'/'VOLT'.")
        # Initialize the buffer
        self.instrument.write(":TRAC:CLE")  # Clear the buffer
        self.instrument.write(f":TRAC:POIN {num_points}")
        self.instrument.write(":TRAC:FEED:CONT NEXT")
        # Perform the sweep
        self.instrument.write(":INIT")
        self.instrument.write(":FORM:ELEM READ")
        return self.instrument.query_ascii_values(":TRAC:DATA?")
# Example usage
controller = Keithley2400Controller()
controller.connect()
sweep_data = controller.iv_sweep(start=0, stop=5, step=0.1, source_type='VOLT', measure_type='CURR')
print(sweep_data)
\end{lstlisting}
\textcolor{darkblue}{In this method:}
\begin{itemize}
    \item \textcolor{darkblue}{\texttt{start}, \texttt{stop}, and \texttt{step} define the sweep range and step size.}
    \item \textcolor{darkblue}{\texttt{source\char`_type} and \texttt{measure\char`_type} determine whether you're sourcing voltage and measuring current, or sourcing current and measuring voltage.}
    \item \textcolor{darkblue}{The buffer is set up to store the number of points calculated from the sweep parameters.}
    \item \textcolor{darkblue}{The method returns the data from the sweep, which you can then process or plot as needed.}
\end{itemize}
\end{mdframed}
\caption{User-ChatGPT interaction to implement a current-voltage (IV) sweep method.}
\label{s3}
\end{figure*}

\clearpage

\end{document}